\documentclass[12pt,preprint]{aastex}

\def\spose#1{\hbox to 0pt{#1\hss}} 
\def\simlt{\mathrel{\spose{\lower 3pt\hbox{$\mathchar"218$}}      
\raise 2.0pt\hbox{$\mathchar"13C$}}} 
\def\simgt{\mathrel{\spose{\lower 3pt\hbox{$\mathchar"218$}}      
\raise 2.0pt\hbox{$\mathchar"13E$}}} 
 \def\ie{{\rm i.e.~}} 
\def\etal{{\rm et~al.~}} 
\def\mic{{~$\mu$m\ }}
\def\mice{{~$\mu$m}}

\lefthead{Mould et al} 
\righthead{M31 Spitzer} 
\begin{document}
\title{A Point Source Survey of M31 with the Spitzer Space Telescope}

\author{Jeremy Mould} 
\affil{School  of Physics, University of Melbourne, Vic 3010, Australia} 
\authoraddr{E-mail: jmould@ph.unimelb.edu.au} 

\author{Pauline Barmby} 
\affil{Department of Physics \& Astronomy, University of Western Ontario, 1151 Richmond St, London ON N6A 3K7, Canada}

\author{Karl Gordon} 
\affil{Space Telescope Science Institute, 3700 San Martin Drive, Baltimore, MD 21218} 
\authoraddr{E-mail: kgordon@stsci.edu} 

\author{S.P. Willner \& M.L.N. Ashby}
\affil{Harvard Smithsonian Center for Astrophysics, 60 Garden St, Cambridge, MA 02138}
\authoraddr{E-mail: willner@cfa.harvard.edu, mashby@cfa.harvard.edu}

\author{R.D. Gehrz, Roberta Humphreys, \& Charles E. Woodward}
\affil{Department of Astronomy, University of Minnesota, 116 Church St SE,
Minneapolis, MN 55455} 
\authoraddr{E-mail: gehrz@astro.umn.edu, roberta@astro.umn.edu, chelsea@astro.umn.edu}

\begin{abstract}
We explore the stellar population of M31 in a Spitzer Space Telescope survey utilizing
IRAC and MIPS observations. Red supergiants are the brightest objects seen
in the infrared; they are
a prominent evolutionary phase. Due to their circumstellar envelopes, many of these radiate the bulk
of their luminosity at IRAC wavelengths and do not stand out in the near
infrared or optically. Going fainter, we see large numbers of  luminous asymptotic
giant branch stars, many of which are known long period variables. Relative to
M33 the AGB carbon star population of M31 appears sparse, but this needs
to be spectroscopically confirmed.
\end{abstract}


\section{Introduction}
The disk and bulge of M31 were first resolved into stars by \cite{hub36}
and \cite{baa44}, respectively. Since then, many optical surveys have been carried out,
including those of \cite{hs53}, \cite{bs63}, \cite{hod76}, \cite{ber88}, \cite{mag92}, \cite{moc01}
and \cite{mas06}. More recent papers that included spectra and optical and near-infrared
photometry of individual stars are \cite{h79}, \cite{h88}, \cite{hmf90}, \cite{h94} and \cite{mas98}.
Near-infrared point-source photometry of M31 
has been hampered by the lack of large-format detectors; for example, the recent work of \cite{ols06}
covers only a small portion of M31. The 2MASS Large Galaxy Atlas \citep{jar03}
presented integrated photometry of M31, but no galaxy-wide assessment of the M31 point
source population has been published.

Spitzer Space Telescope characterization of the point source population of other Local Group galaxies
includes the Large Magellanic Cloud \citep{blu06}, M33 \citep{ver07,mcq07},
and NGC 6822 \citep{can07}. Other Local Group dwarfs studied include IC1613 and the WLM galaxy \citep{jaa07,jab07,jac06}.
The Andromeda galaxy, M31, presents a unique opportunity to study stellar evolution at a chemical composition
similar to that of the Milky Way. The red giant branch was
resolved in V \& I by \cite{mk86},  and the brightest evolved stars have been studied by \cite{mas98}.
However, an infrared survey of point sources in M31 has awaited the arrival of the Spitzer Space Telescope 
\citep{wer04,geh07} and 2MASS \citep{jar03}.

In the mid-infrared, integrated measurements of M31 have been made by IRAS 
\citep{hab84,xuh96},
COBE/DIRBE \citep{ode98},  MSX \citep{kra02}
and {\it Spitzer} \citep{gor06,bar06}. With the aim of reaching stars
with the most significant mass loss rates this
 paper presents a point source survey with the IRAC \citep{faz04}
and MIPS \citep{rie04}  cameras.
We also utilize the 2MASS survey in order to include shorter wavelengths.
We present extensive tables, giving positions of the sources we identify, to enable follow-up of these observations.
 
\section{IRAC Survey}
The IRAC data and reduction are described in \cite{bar06}. Briefly, the central
$1\fdg6 \times 0\fdg4$ of
M31 was imaged to a depth of six 12-second frames per position, and the 
outer area to a size of $3\fdg7 \times 1\fdg6$ by four 30-second frames per position.
The survey was part of the Spitzer Cycle 3 program (Program Identification 3126).
The data analyzed here were produced by version 14 of the Spitzer Science Center (SSC) pipeline;
compared to the version 11 and 12 data used by \cite{bar06}, the newer version is
expected to have more accurate pointing refinement for the individual frames.
The 3000 individual frames per channel were combined with the MOPEX software
to create mosaics with a pixel scale of 0\farcs86, sub-sampling the native IRAC
pixel scale by a factor of $\sqrt{2}$ to better sample the point spread function 
(FWHM $\sim1\farcs9$). 

The saturation limits given in the IRAC Data Handbook \footnote{http://ssc.spitzer.caltech.edu/irac/sat.html}
correspond to Vega magnitudes of
$(9.9, 9.3, 6.7, 6.8)$ for 12-second frames and $(10.8, 10.4, 7.7, 7.8)$ 
for 30-second frames for the Spitzer 3.6, 4.5, 5.8, 8 \micron~ bands respectively.
The sensitivity can be estimated using the SSC Performance 
Estimation Tool (http://ssc.spitzer.caltech.edu/irac/sens.html);
 assuming a medium background, the $5\sigma$ point source 
detection limits correspond to Vega magnitudes of
$(19.0, 18.0, 15.6, 14.9)$ for six 12-second frames and $(19.6, 18.5, 16.0, 15.2)$ 
for four 30-second frames. However, point source photometry in the vicinity of M31 
is affected by crowding and at longer wavelengths by the diffuse ISM, and these values are presented as a guide only.

The IRAC flat-fields are derived from observations of the zodiacal light and 
observations of calibration stars to correct for the fact that the large-scale sensitivity
pattern varies with spectral shape. Using the `photometric correction' images
from the Spitzer Science Center, we constructed correction mosaics and applied
them to the M31 mosaics.
We tested two different corrections for alternately red and blue sources and
found that the effect was small. The magnitude changes for the red source correction are
$(0.02, 0.0, 0.0, 0.0)$ in the sense that point sources become fainter
and the scatter about this mean is $(0.16, 0.04, 0.02, 0.02)$. With the blue source correction
stellar magnitudes in Table 1 become brighter by \\$(-0.11,-0.08,-0.06,-0.04)$.
The scatter about these means is $(0.20, 0.13, 0.13, 0.10)$. Therefore, we have not implemented
a correction of this type, as the effect is small, and the conclusions below are unaffected.

\section{MIPS Survey}

The MIPS resolution ranges from 6 arcsec at 24\mic to 40 arcsec at 160\mice.
\cite{gor06}  described the MIPS program surveying M31,
fit a stellar population plus dust grain model to the spectral energy distribution,
and discussed the structure of M31 in terms of its interactions with M32 and NGC 205. 
Point sources were extracted from the 7 individual MIPS 24\mic
AOR mosaics that together cover approximately $1\times3$ degrees centered on
M31.  

The point source extraction was done using the PSF-fitting
program StarFinder \citep{dio00} using a MIPS 24\micron~ model PSF
appropriate for stars \citep{eng07}.  This method of
extracting point source photometry has been extensively tested on LMC
Spitzer observations \citep{mei06} and found to produce high
quality point source catalogs at 24\mice.

\section{Stellar Photometry}

Point spread functions (PSFs) were taken from the MOPEX\footnote{http://ssc.spitzer.caltech.edu/irac/postbcd/mopex.html}
analysis release by the Spitzer Science Center.
Sixteen realizations of the PSF differing in pixel phase were constructed for each of the IRAC channels 
at the survey pixel scale and combined with DAOPHOT \citep{ste87}. 
Two passes were made with the ALLSTAR PSF-fitting software
for each of the four channels, yielding over a quarter of a million point sources at 3.6\mic ~with 
a detection threshold of 2.5$\sigma$. At 8.0\mic ~we detected
160,000 sources with the same threshold. Source lists were merged with the DAOMASTER program \citep{ste92}, using a 1 pixel matching radius. 
Calibrated following \cite{rea05}, and using aperture corrections based on the PSFs, Figure 1
is the IRAC 2-color diagram for these objects. It resembles the equivalent diagram for the Large
Magellanic Cloud \citep{mei06}. Dusty evolved stars are offset from the origin in Figure 1.
Objects with [5.8]--[8] $>$ 1 are young stellar objects and show some concentration to the ring
of star formation in M31 \citep{hab84}, as seen in Figure 2. Figure 3 compares
the luminosity functions in the inner and outer parts of the galaxy.

An estimate of the
foreground star and background galaxy contamination may be obtained by comparison with a survey such as SWIRE 
(Lonsdale \etal 2004). Its lowest galactic latitude
field is ELAIS-N2; $b~=~+42^\circ$, compared to M31's $b~=~-21.2^\circ$. Its IRAC 2-color diagram is reproduced in 
Figure 4. The IRAC catalog in this field covers $\sim 4.2$~deg$^2$.  
To a magnitude limit of [3.6] = 17, there are 40,953 objects in the ELAIS-N2 catalog;
scaled to the IRAC M31 area (3.08~deg$^2$), this corresponds to 29,895 objects, a tenth of the total to 17th mag in M31. 
The color distribution of the SWIRE sources
extends further to the red in [5.8]--[8.0] than the IRAC M31 sources, but the M31 sources' color extends further 
to the red in [3.6]--[4.5]. The brighter M31 sources
also extend further to the red in [3.6]--[8.0]. All of these properties of the color distributions 
are consistent with the M31 sources being dominated by stellar objects rather than
galaxies, particularly at the faint end.

Figure 5 is a color magnitude diagram (CMD) depicting massive star evolution in M31. Most of the stars
fall on a strong vertical feature. These are main sequence and supergiant stars, together with significant 
contamination from foreground stars. The horizontal feature at 
[3.6] = 10 mag shows the growth of circumstellar envelopes radiating at 300--1000K. To enable
follow-up of these objects, magnitudes for
[3.6] $<$ 11 mag are recorded in Table 1. M31 is only a 29\%
overdensity for 9 $<$  K $<$ 11 mag in the 2MASS point source catalog. (We
determined this by catalog searches in the four cardinal directions around
M31.)
We therefore expect 1/1.29, \ie 77 $\pm$ 6 \%, of the entries in Table 1 to be foreground.
Figure 6 shows the SWIRE control field. Scaling the 268 ELAIS-N2 catalog
with [3.6] $<$ 11 mag to the area of our M31 field, we expect 196 such objects.
There are 426 stars in the M31 field with [3.6] $<$ 11 and measured [8.0] mags.
If we assume that foreground objects scale as cosec($b$), the estimate of 196 foreground
stars becomes 362, which provides an independent estimate of the percentage
foreground contamination of 85 $\pm$ 7\% . That is a 2$\sigma$ detection of evolved
stars in M31 with [3.6] $<$ 11 mag.

The maximum
luminosity [3.6] = 9.5 mag corresponds to $M_{\rm bol}\approx -11.9$ 
according to the supergiant bolometric corrections of \cite{joh66} for a distance of $m-M = 24.4$ mag
\citep{hut95}. This is a conservative estimate, as $[3.6] = 9.5$ mag
exceeds the saturation limits noted in $\S$2. Such high luminosities (1--2.5 $\times~10^6~L_\odot$) are reached in the evolution of 
120 M$_\odot$ stars \citep{chi86}. 

The tip of the asymptotic giant branch (AGB) \citep{woo83}
at M31 lies at $[3.6] > 14$ mag: all the evolved stars shown in Figure 5 are supergiants.
The positions of $\eta$ Car and VY CMa are represented on the diagram, moved to the distance of M31.
Data for these objects are from \cite{smi01}, \cite{pol99}, and \cite{whi94}.
Most of the stars with $[3.6] \approx 10$ are bluer than these objects: they are presumably
in an earlier evolutionary phase of thickening circumstellar envelopes
\footnote{RY Scuti is another case in point \citep{geh01}. It has [3.8] -- [8.9] $\approx$ 2.9 mag, but is considerably fainter than $\eta$ Car and VY CMa at 3.6\micron.}.
Figure 7 extends deeper into the CMD for M31. Although photometric errors become significant for
[3.6] $\sim$~ 17, we see a large number of AGB stars, stretching from the tip at 14.4 to the detection
limit at 17 mag. Figure 8 is the corresponding SWIRE control field.

We carried out completeness tests by adding artificial stars to the [3.6] image. Completeness falls to
50\% at [3.6] = 15.5 mag and drops to almost zero at 16.5 mag.

We cross-correlated our catalog with that of \cite{mas06}. Over 87,000 matches were found within 
1.6 arcsec, but only a handful of these have been spectroscopically classified. Indeed, only 11 of these
matches are included in Table 1, and these are (with two exceptions) 21st and 22nd V mag stars. Luminous
Blue Variables are also known in M31 \citep{sz96}, but these have K magnitudes between 13.9 and 17.6.
Clearly, there are many stars representing the latest stages of evolution in the stellar population of M31 
that are most prominent in the infrared.

We can also compare the CMD of M31 with that of M33 in Figure 9. \cite{mcq07}
plot the IRAC [3.6] -- [8] color distribution by [3.6] magnitude in
their figure 19. In M33 M$_{3.6} <$ --10 consists of AGB stars without dust;
the other two histograms are carbon stars with dust. We consider the carbon
star frequency of M31 in the next section.

\section{Spectral Energy Distributions}
We cross-correlated our source catalog with the 2MASS Sky Survey \citep{skr06}
using the OpenSkyQuery facility (http://openskyquery.net/Sky/skysite )
of the National Virtual Observatory. Some 923 matches were found. 
To go deeper in the near infrared, we downloaded a catalog from the ancillary
data products of the 2MASS Extended Mission. This derives from a set of
special observations made at the end of 2MASS survey operations that used
exposure times 6 times longer than the main survey. These measurements
reach $\sim$1 mag deeper than the survey. We found 3110 matches with the 6 $\times$ catalog within a 1.6 arcsec radius.

Figure 10 is a CMD plotted from these matches. Following the deep JHK photometry
of M33 by \cite{ci08}, the strong vertical feature with J--K $<$ 1 and K $<$ 14.5 is foreground, mixed with M supergiants. The broader vertical feature with K $>$
14.5 are M type AGB stars and stars with J--K $>$ 1.36\footnote{As M31 is more metal rich than M33, this cutoff is probably redder in M31. E(J--K) is only 0.01 mag larger in M31 than M33 \citep{sch98}.}  are candidate carbon stars. The principal difference between M31 and M33, besides the fact that the
UKIRT M33 photometry goes deeper, is that the branch of carbon stars at
K = 16 mag (in
Figure 2) of \cite{ci08} is mostly absent, replaced by a dozen or so stars in M31
at K $\approx$ 14 mag. The product of third dredge up, carbon stars are harder to make in metal rich envelopes \citep{rv81}.

\cite{blu06} have classified stars in the SAGE survey of the Large Magellanic
Cloud \citep{mei06} according to their position in the [3.6], J-[3.6] CMD.
Figure 11 reproduces this classification for an LMC distance modulus of 18.5
\citep{mo2000}. Supergiants are separated in color from AGB M stars, which
are in turn separated from AGB C stars. In the LMC this separation is clearly
made into distinct concentrations in the Hess diagram, in which the RGB
tip, the oxygen-rich and carbon-rich AGB branches are clearly visible. 
In M31 the IRAC/2MASS CMD is sparser and noisier, and additional dredge up 
may be required to overcome additional metallicity and
 make carbon stars. Classification based on Figure 10 may be indicative
rather than exact, until spectroscopy can be obtained at 14 $<$ [3.6] $<$ 16
and J--K $\approx$ 2.

Matching with 6$\times$ 2MASS allowed spectral energy
distributions (SED) to be plotted from 1--8~\micron. Foreground stars will tend to show photospheric SEDs.
For those stars showing a mid-infrared excess (defined as a peak at 3.6~\micron  ~or beyond),
we display SEDs in Figure 12 and magnitudes in Table 2. Three broad classes of
sources are seen in Figure 12: (i) those that peak at 3.6~\micron, (ii) sources
with a longer wavelength excess, and (iii) relatively flat spectrum sources. The 
first class are consistent with rather warm circumstellar envelopes ($\sim$1000K);
in class (ii) the envelope is cooler ($\simlt$500K); class (iii) would be consistent with compact
HII regions, supernovae remnants, or planetary nebulae, especially if the
approximately flat spectrum extends to 24~\micron~ (Verley \etal 2007). Omitted
from Figure 12 are cases where DAOMASTER failed to match a star at an individual wavelength. 
There are approximately 20 such cases for which spurious peaks appear in the SED.

Bolometric corrections are not tabulated for stars as red as some of these objects. Integrating these
SEDs from 1 -- 10~\micron, we obtain luminosities from 1 -- 20 $\times$ 10$^4$ L$_\odot$ for those
that are members of M31.  The majority of these stars are AGB stars with 
circumstellar envelopes; there are 12 that are nominally supergiants (
M$_{\rm bol}~<$ --7), mostly with warm circumstellar envelopes. 
We cannot rule out that some of these stars may be late type dwarfs
in the foreground.

\section{Long Period Variables}

The catalog was also cross-correlated with the long period variable (LPV) catalog of \cite{mou04}
which covers about half of the disk of M31. 
We found 831 sources within 2.5 arcsec of LPVs, triple the number that
would have been expected by chance. The matches appear in Table 3
 and comprise over 40\% of the LPV catalog.  Figure 13 is the IRAC 2-color diagram for LPVs and Figure 14 is a color magnitude diagram.

The distribution of points in Figure 14 is bounded by the limits of Mould, Saha \& Hughes's photometry
at J $\approx$ 18.5 and the present photometry [3.6] $\approx$ 17 mag.
With $\langle K\rangle = 16.38$ mag, these LPVs have $M_{\rm bol} \approx -5$ and are AGB stars. This contrasts
with the LMC and M33 which show significant numbers of supergiant LPVs with $M_{\rm bol}< -7$ in addition to
their AGB population \citep{woo83,mou90}.

\section{MIPS sources}
Over 25,000 sources from 1 Jy to 0.2 mJy were extracted from the MIPS 24~\micron ~image. Sources that are bright
at 24~\micron~ include HII regions, supernova remnants, and planetary nebulae (Verley \etal 2007).
Nevertheless, cross-correlation with the brightest six magnitudes of the IRAC stellar
catalog yielded 599 matches within 1 arcsec, and these are found in Table 4. Sixty matches would have been expected by chance.
Figure 15 shows that
that many of these matches fall on the dusty stars side of the IRAC-MIPS 2-color diagram,
rather than the young stellar objects side (Meixner \etal 2006).  J004258.15+410731.3 is 667 day period
LPV 33133.  J004031.89+410624.7 is 785 day period LPV 16094.
 J004029.18+404453.8 is 575 day period LPV 14023. These are
3 of the 20 longest periods in the LPV catalog. Their SEDs are shown in Figure 16. These can be compared
with some of the brighter MIPS sources in Figure 17. These extreme period LPVs appear to show two components,
a photospheric component, and a dust shell that is relatively cool compared with those illustrated in $\S$5.
 
The luminosity function of 24~\micron~ sources is illustrated in Figure 18. Brighter sources tend to be contained
within M31's ring of star formation and are more likely to be compact HII
regions or young stellar objects.

\section{Conclusions}
Our Spitzer Space Telescope survey reveals a rich population of evolved stars in M31. Although there is strong foreground contamination, supergiants are present as bright as
$1 - 2.5 \times 10^6$~$L_\odot$. These evolve into stars with extensive circumstellar envelopes like $\eta$~Carinae
and VY CMa in the Milky Way. A number of the brightest supergiants at 3.6\mic are so red that they
will be missed in optical surveys.
We also resolve stars below the tip of the AGB to $M_{\rm bol} = -5$~mag. Many of these are LPVs, which also have dusty envelopes. Photometric candidates for
AGB carbon stars are much fewer in M31 than M33, but spectroscopy is required
to quantify this.

\acknowledgements
This work is based on observations made with the Spitzer Space Telescope, 
which is operated by the Jet Propulsion Laboratory, California Institute of Technology under a contract with NASA. 
Support for this work was provided by NASA through an award issued by JPL/Caltech.
This work makes use of 2MASS data products, a joint project of the University of Massachusetts and IPAC/Caltech, funded by NASA and NSF. 
In addition to DAOPHOT, this research has made use of IRAF, which is distributed by NOAO. NOAO is
operated by AURA under a cooperative agreement with NSF.


\begin{deluxetable} {rrrrrrr}
\tablecaption{\bf Photometry of the infrared brightest stars}\label{tab1}
\tablehead{
\colhead{\#} & \colhead{RA (2000)} & \colhead{ Dec } & \colhead{[3.6]} & \colhead{[4.5]} & \colhead{[5.8]} & \colhead{[8]}  }
\startdata
J003648.82+395625.7&  9.20341& 39.94046&  10.87  7&  10.82  5&  10.80  2&  10.71  3 \nl
J003659.17+400336.6&  9.24653& 40.06018&   9.67 11&   9.36  2&   9.23 10&   9.00  1 \nl
J003700.47+395031.8&  9.25196& 39.84217&  10.26  5&  10.07  3&  10.02  5&   9.88  2 \nl
J003700.75+401914.0&  9.25311& 40.32055&   9.99 12&   9.76  5&   9.66  5&   9.60  2 \nl
J003700.97+395619.2&  9.25404& 39.93867&  10.18 18&  10.79 21&  12.12 40&  11.19 30 \nl
J003708.35+403034.8&  9.28479& 40.50966&  10.16  7&   9.24 10&   8.52  1&   8.37  1 \nl
J003708.38+403032.4&  9.28490& 40.50901&  10.33  7&          &  11.03 11&  11.24 10 \nl
J003708.54+394701.7&  9.28557& 39.78380&  10.82  7&  10.92  5&  10.93  7&  10.79  5 \nl
J003709.07+401450.0&  9.28779& 40.24722&  10.07 10&   9.93  5&   9.87  3&   9.77  1 \nl
J003709.53+403350.1&  9.28970& 40.56393&  10.98  5&  10.81  7&  10.91  2&  10.87  7 \nl
J003716.53+395705.7&  9.31886& 39.95158&   9.78  7&   9.58  5&   9.61  1&   9.43  1 \nl
\ldots  & \ldots & \dots & \ldots & \ldots & \ldots & \ldots \nl
\enddata
\tablecomments{Magnitudes are given together with an uncertainty in hundredths of a magnitude. Table 1 is published in its entirety in the electronic edition
of the $Astrophysical~ Journal$. A portion is shown here for guidance regarding
its form and content.}
\end{deluxetable}

\begin{deluxetable} {rrrrrrrrrr}
\tablecaption{\bf 2MASS 6$\times$ matches}\label{tab2}

\tablehead{
\colhead{\#} & \colhead{RA (2000)}&\colhead{Dec} &\colhead{ J}  & \colhead{ H }&\colhead{K}& \colhead{[3.6]}&
\colhead{[4.5]}& \colhead{[5.8]} & \colhead{[8.0]}}
\startdata
     52&   9.62906&  40.30904&  17.17&  16.64&  17.16&  11.50&  11.59&  11.38&  11.35 \nl
     24&   9.67014&  40.47509&  17.57&  16.80&  16.16&  13.04&  12.98&  12.95&  12.85 \nl
      3&   9.72386&  40.49491&  18.25&  16.67&  16.22&  14.01&  13.28&  12.79&  12.28 \nl
     43&   9.72643&  40.62497&  17.69&  16.62&  16.30&  12.03&  12.10&  11.74&  12.17 \nl
     49&  10.07809&  40.60300&  17.69&  16.36&  16.26&  13.72&  12.52&  11.37&  10.41 \nl
     15&  10.08307&  40.56743&  17.65&  16.57&  16.43&  14.69&  13.47&  12.73&  11.82 \nl
      6&  10.08431&  40.67662&  17.23&  16.34&  15.86&  14.87&  13.70&  12.43&  11.57 \nl
     44&  10.12849&  40.70961&  16.99&  15.78&  15.59&  14.84&  14.19&  12.03&  10.87 \nl
   3075&  10.13519&  40.65016&  16.37&  16.20&  15.32&  15.18&  14.90&  11.97&  10.16 \nl
   2968&  10.29796&  41.13960&  18.31&  16.71&  16.26&  14.88&  14.34&  13.41&  12.57 \nl
   2947&  10.33984&  40.83079&  17.83&  16.77&  16.26&  14.45&  14.04&  11.49&   9.66 \nl
\ldots & \ldots & \ldots & \ldots & \ldots & \ldots & \ldots & \ldots & \ldots &\ldots \nl 

\enddata
\tablecomments{Table 2 is published in its entirety in the electronic edition
of the $Astrophysical~ Journal$. A portion is shown here for guidance regarding
its form and content.}
\end{deluxetable}

\begin{deluxetable} {rrrrrrrr}
\tablecaption{\bf LPV matches}\label{tab3}
\tablehead{
\colhead{MSH(2004) \#} &\colhead{ J}  & \colhead{ H }& \colhead{K}& \colhead{[3.6]}&
\colhead{[4.5]}& \colhead{[5.8]} & \colhead{[8]}}
\startdata
     11009&  18.05&  16.84&  16.38&  16.03&  15.93&       &        \\
     12008&  17.12&  16.07&  16.03&  16.23&       &       &        \\
     12022&  17.63&  17.40&       &  16.17&       &       &        \\
     12026&       &  17.42&       &  16.45&       &       &        \\
     12034&       &  17.30&  17.20&  16.68&       &       &        \\
     12040&  17.90&  17.03&  16.74&  15.84&  15.60&       &        \\
     12041&  18.43&  17.08&  17.54&  15.80&  15.98&       &        \\
     12042&  17.59&  17.03&  17.47&  16.00&  14.99&  14.36&  13.10 \\
     12052&       &  17.97&       &  16.30&       &       &        \\
     12080&  18.58&       &       &  16.19&       &       &        \\
     12090&  18.02&  17.43&  17.82&  15.42&       &       &        \\
     12092&       &  17.82&  17.80&  16.59&       &       &        \\
     13002&  18.53&  17.76&  17.16&  16.04&  16.21&       &        \\
     13012&  17.98&  17.50&  16.98&  15.42&  15.29&  14.59&        \\
     13013&  17.90&  17.26&  16.95&  15.70&  16.11&       &        \\
     13021&  16.15&  15.92&  14.90&  16.52&       &       &        \\
 \ldots   & \ldots& \ldots& \ldots& \ldots& \ldots & \ldots &\ldots \\
\enddata
\tablecomments{Table 3 is published in its entirety in the electronic edition
of the $Astrophysical~ Journal$. A portion is shown here for guidance regarding
its form and content.}
\end{deluxetable}

\clearpage
\begin{deluxetable} {rrrrrrrr}
\tablecaption{\bf MIPS 24$\mu$ sources}\label{tab4}
\tablehead{
\colhead{SSTM1M311 \#} & \colhead{RA (2000)} & \colhead{ Dec }&\colhead{[3.6]}&\colhead{ [4.5]}
&\colhead{  [5.8]}& \colhead{  [8]} &\colhead{[24]} }
\startdata
   J003921.35+402142.2 &  9.83898& 40.36174&  13.98&  13.84&  13.67&  13.65&   9.50 \nl
   J004004.56+403844.1 & 10.01900& 40.64558&  14.62&  14.55&  13.98&  13.51&  10.42 \nl
   J004014.35+404836.3 & 10.05983& 40.81009&  15.12&  14.55&  13.83&  13.46&  10.59 \nl
   J004031.52+404126.6 & 10.13134& 40.69073&  15.86&  15.49&  12.64&  10.75&   5.63 \nl
   J004032.94+403901.7 & 10.13728& 40.65049&  14.89&  14.82&  12.36&  10.65&   8.32 \nl
   J004043.74+404326.6 & 10.18226& 40.72406&  15.51&  15.20&  14.35&  12.88&  10.70 \nl
   J004046.28+403324.6 & 10.19284& 40.55685&  16.04&  15.03&  14.19&  12.96&   9.93 \nl
   J004054.09+404606.5 & 10.22541& 40.76847&  15.68&  15.58&  14.39&  12.43&  10.11 \nl
   J004055.89+404644.3 & 10.23291& 40.77898&  16.03&  15.36&       &  12.57&   9.16 \nl
   J004057.04+404935.8 & 10.23771& 40.82663&  15.75&  15.22&  14.30&  13.92&   9.65 \nl
   J004059.78+403530.7 & 10.24912& 40.59188&  11.39&  11.35&  11.14&  10.87&  10.35 \nl
   J004102.24+410432.1 & 10.25936& 41.07561&  15.73&  15.51&  12.80&  10.77&   8.31 \nl
   J004104.65+405428.4 & 10.26940& 40.90791&  12.43&  12.53&  12.28&  12.25&  11.34 \nl
 \ldots   & \ldots& \ldots& \ldots& \ldots& \ldots & \ldots &\ldots \\
\enddata
\tablecomments{Table 4 is published in its entirety in the electronic edition
of the $Astrophysical~ Journal$. A portion is shown here for guidance regarding
its form and content.}
\end{deluxetable}
\clearpage
\figcaption{IRAC 2-color diagram for sources in M31. The magnitudes are
standard IRAC Vega magnitudes.}

\figcaption{Distribution of sources with measured [8.0] magnitudes in M31. Sources with [5.8] -- [8.0]
$>$ 1 are denoted by red triangles. The galaxy in the western ``fin" of this plot is NGC 205. The northern major axis of M31 is to the left.}

\figcaption{Source counts as a function of magnitude. The shaded area
refers to the inner (deprojected) 9 kpc of M31. The unshaded area shows
the full source counts.}

\figcaption{IRAC 2-color comparison diagram for the ELAIS-N2 SWIRE field.
Sources brighter than [3.6] = 17 mag are shown for comparison with Figure 1.}

\figcaption{IRAC color magnitude diagram for sources in M31. The positions of two Galactic evolved
stars are shown at M31's distance. Also shown is M33 Variable A (Humphreys \etal 2006). The distance
modulus assumed for M33 is 24.52 mag (Lee \etal 2002).}

\figcaption{IRAC color magnitude diagram for the ELAIS-N2 SWIRE control field.}

\figcaption{Deep IRAC color magnitude diagram for sources in M31. This CMD extends to the AGB.
Photometric errors ($\pm 1\sigma$) are displayed at the right of the figure.}

\figcaption{Deep IRAC color magnitude diagram for the ELAIS-N2 SWIRE control field.}

\figcaption{Color distribution as a function of magnitude for comparison
with M33, see \cite{mcq07}. In M33 major peaks in the histograms
indicate AGB stars without dust ([3.6]--[8] $\approx$ 0), carbon stars
with dust (0.5 $\simlt$ [3.6]--[8] $\simlt$ 2.5) and M$_{3.6} <$ --9, and
YSOs ([3.6]--[8] $\simgt$ 4).}

\figcaption{CMD for the matches with 6$\times$ 2MASS.}

\figcaption{CMD following the classification by \cite{blu06} of stars into
supergiants (green), AGB oxygen-rich (blue), and AGB carbon-rich (red).}

\figcaption{Spectral energy distributions for sources in 2MASS with mid-infrared excesses.
The colored numbers refer to Table 2.}

\figcaption{Two color diagram for M31's LPVs.}

\figcaption{CMD for M31's LPVs.}

\figcaption{IRAC-MIPS 2-color diagram for bright sources in M31 from Table 4.
According to Meixner \etal 2006, young stellar objects lie on the
right side of the dashed line.}

\figcaption{Spectral energy distributions for 3 LPVs.
The colored numbers are those of Mould, Saha, \& Hughes (2003).}

\figcaption{Spectral energy distributions for 5 bright 24\micron ~sources in M31.}

\figcaption{Luminosity function for 24\micron ~sources. The red (dashed) histogram is for sources
contained within M31's ring of star formation.}

\clearpage

\begin{figure}
\plotone{f1.eps}
\centerline{f1.eps}
\end{figure}
\clearpage

\begin{figure}
\includegraphics[angle=0,width=0.5\textwidth]{f2.eps}
\centerline{f2.eps}
\end{figure}
\clearpage

\begin{figure}
\plotone{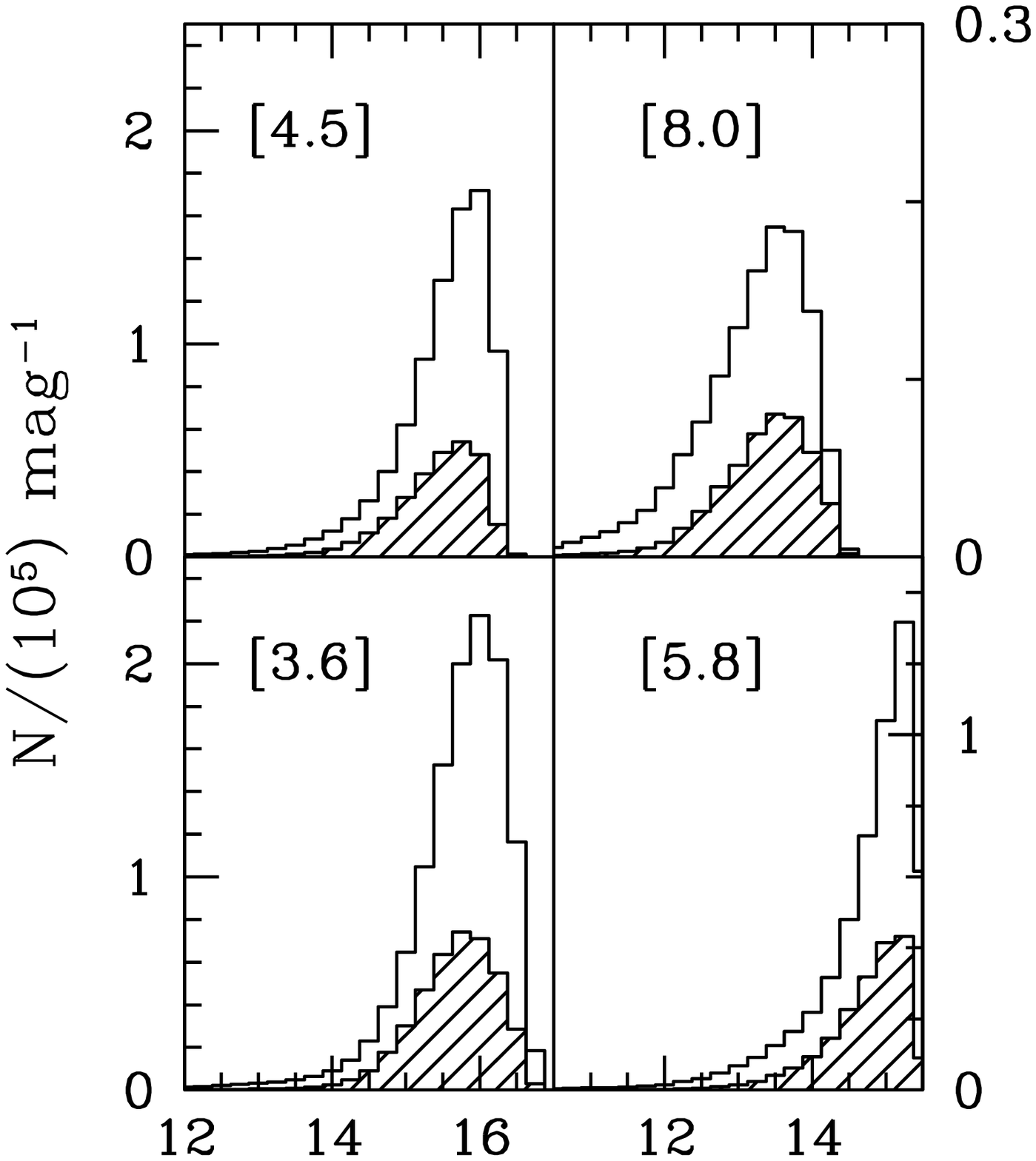}
\centerline{f3.eps}
\end{figure}
\clearpage

\begin{figure}
\plotone{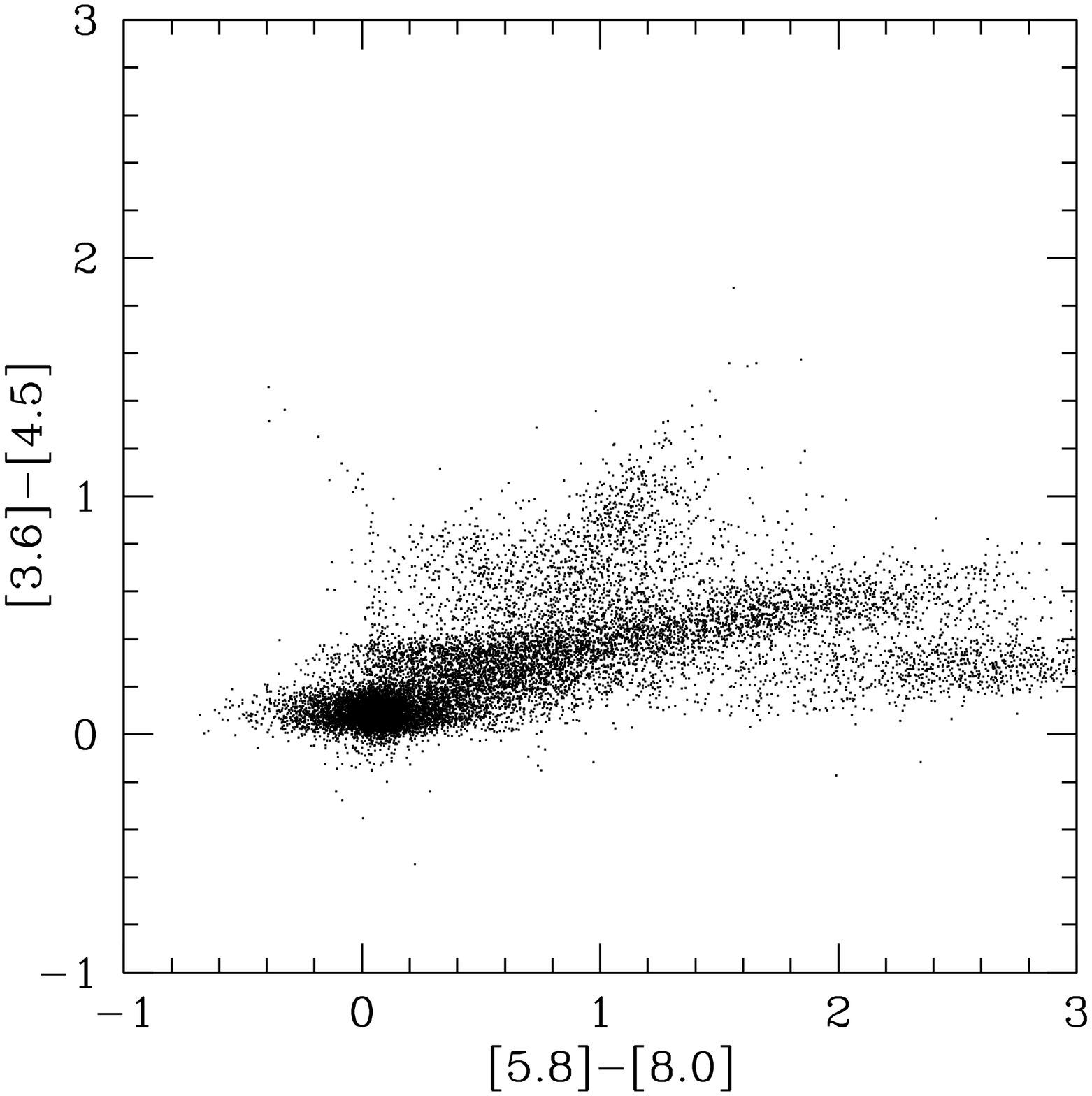}
\centerline{f4.eps}
\end{figure}
\clearpage

\begin{figure}
\includegraphics[angle=0,width=0.8\textwidth]{f5.eps}
\centerline{f5.eps}
\end{figure}
\clearpage

\begin{figure}
\plotone{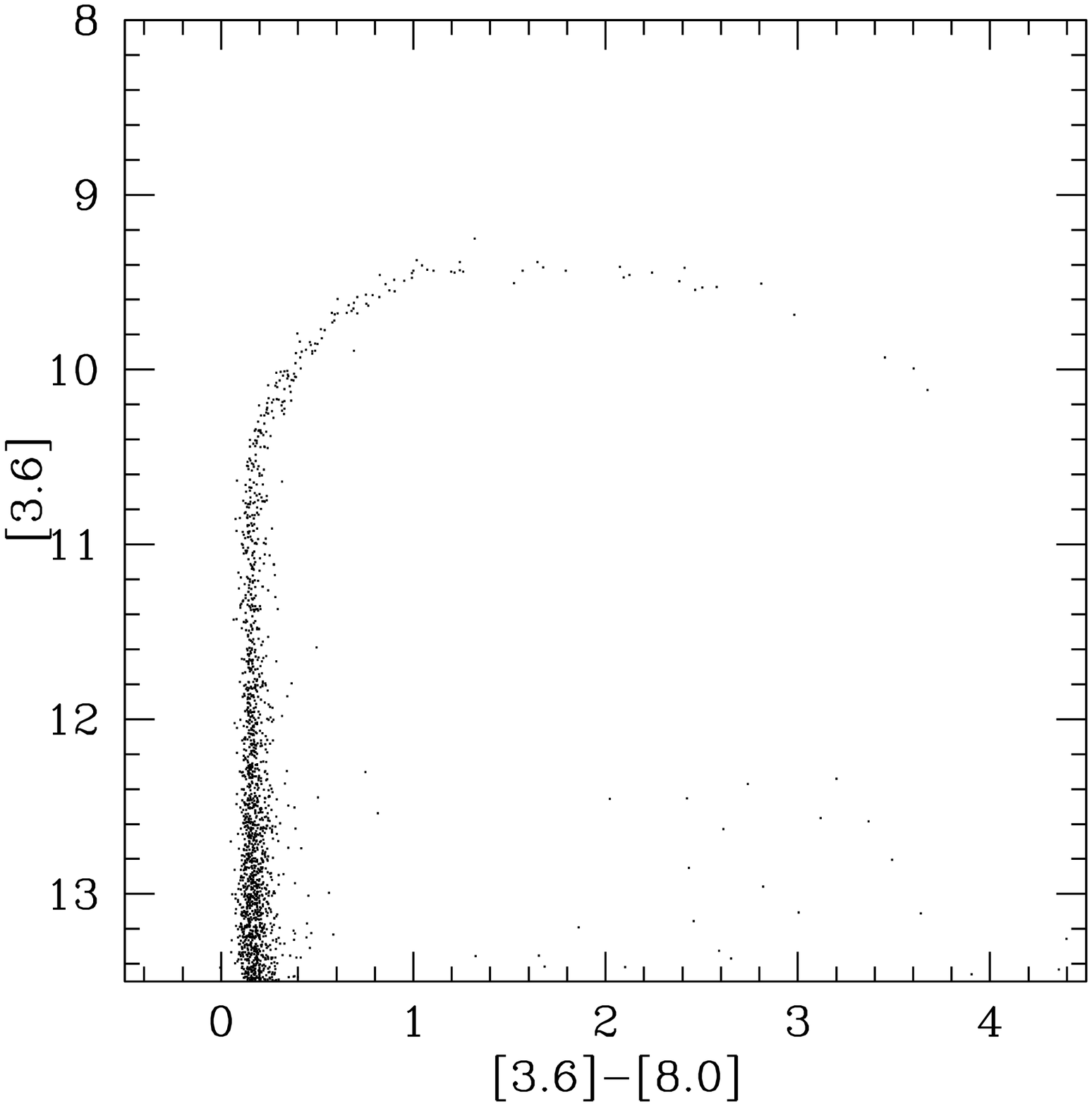}
\centerline{f6.eps}
\end{figure}
\clearpage

\begin{figure}
\includegraphics[angle=0,width=0.8\textwidth]{f7.eps}
\centerline{f7.eps}
\end{figure}
\clearpage

\begin{figure}
\plotone{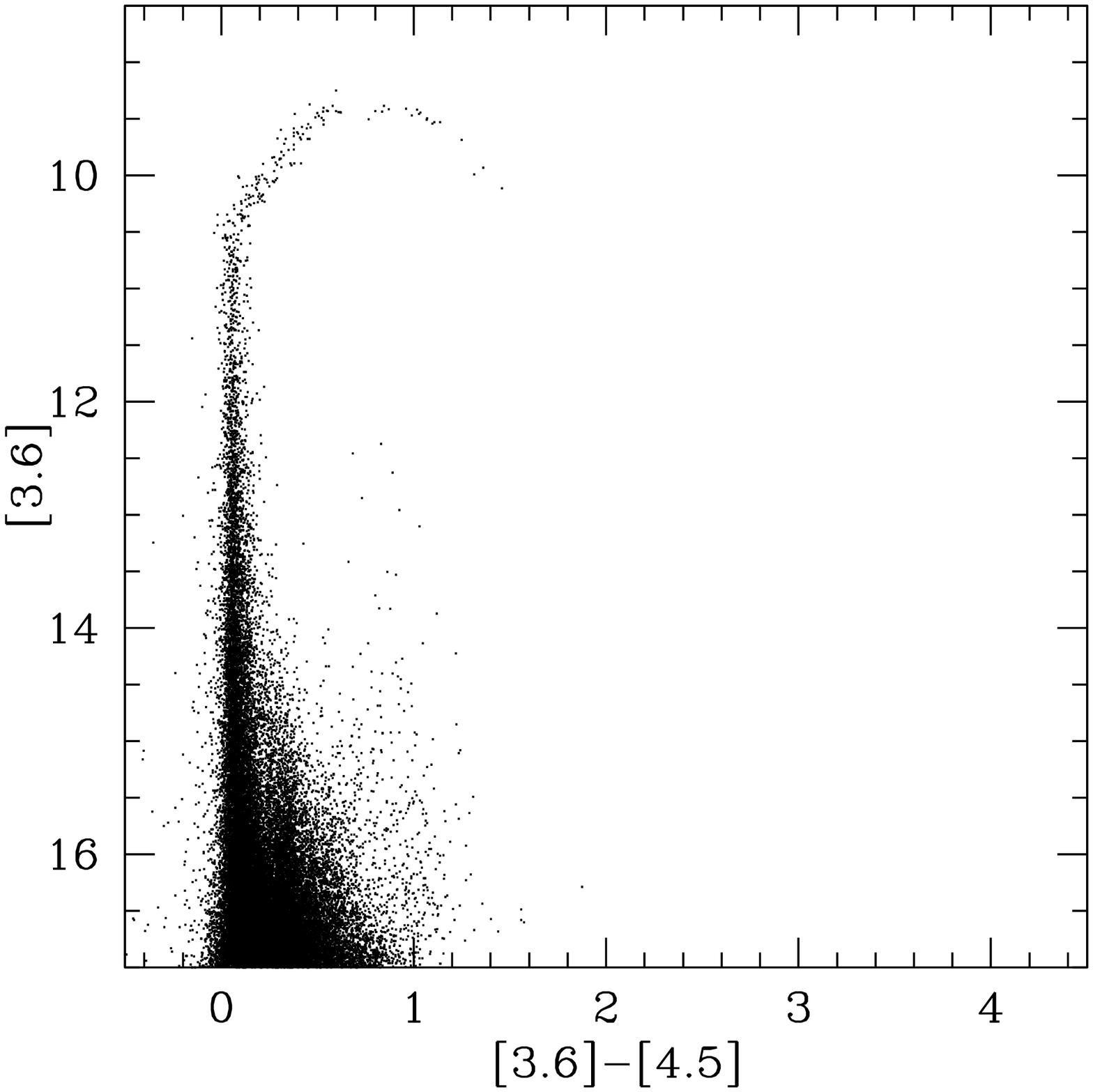}
\centerline{f8.eps}
\end{figure}
\clearpage

\begin{figure}
\plotone{f9.eps}
\centerline{f9.eps}
\end{figure}
\clearpage

\begin{figure}
\plotone{f10.eps}
\centerline{f10.eps}
\end{figure}
\clearpage

\begin{figure}
\includegraphics[angle=-90,width=1.2\textwidth]{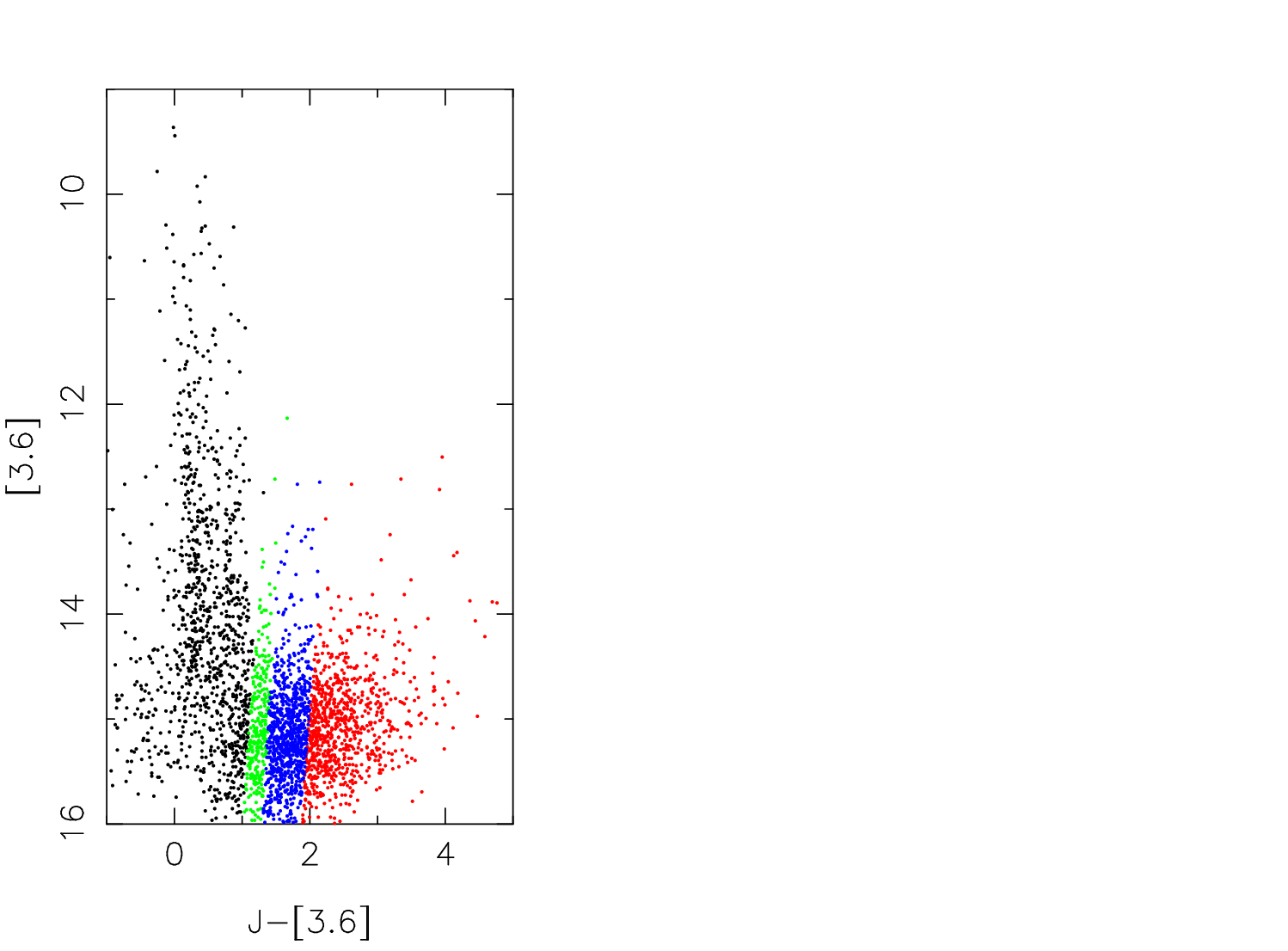}
\centerline{f11.eps}
\end{figure}
\clearpage

\begin{figure}
\includegraphics[angle=-90,width=\textwidth]{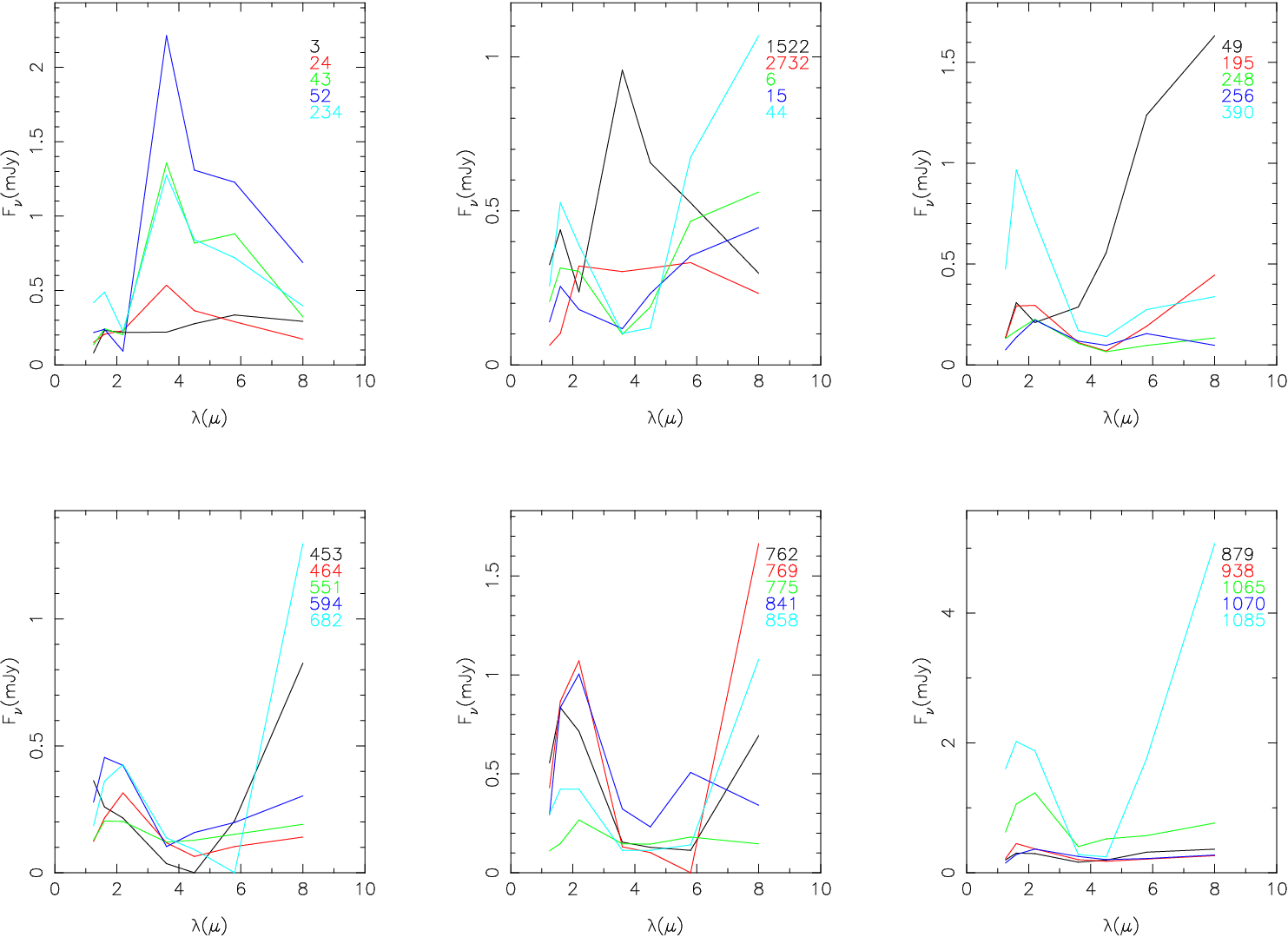}
\centerline{f12a.eps}
\end{figure}
\clearpage

\begin{figure}
\includegraphics[angle=-90,width=\textwidth]{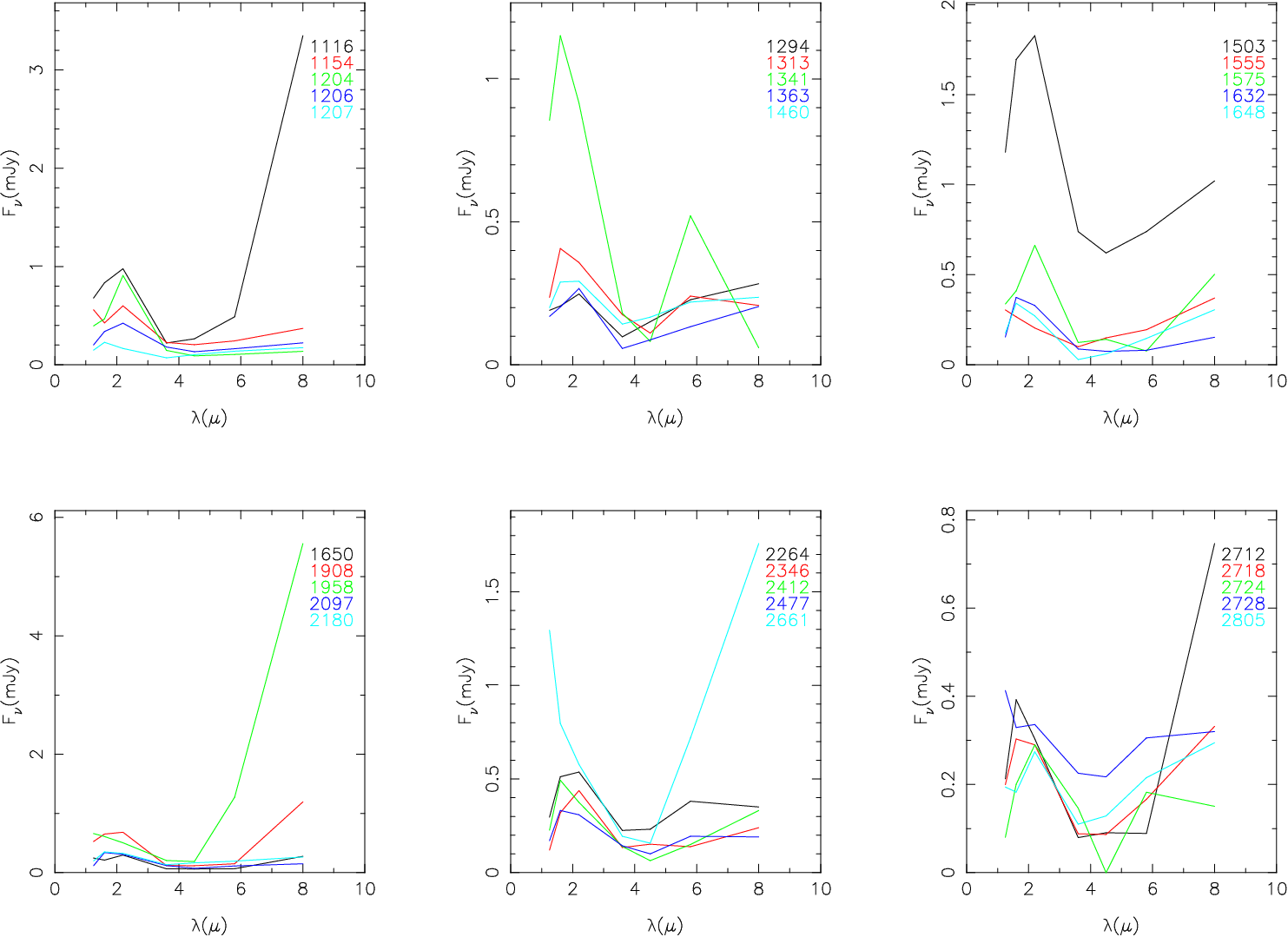}
\centerline{f12b.eps}
\end{figure}
\clearpage

\begin{figure}
\includegraphics[angle=-90,width=\textwidth]{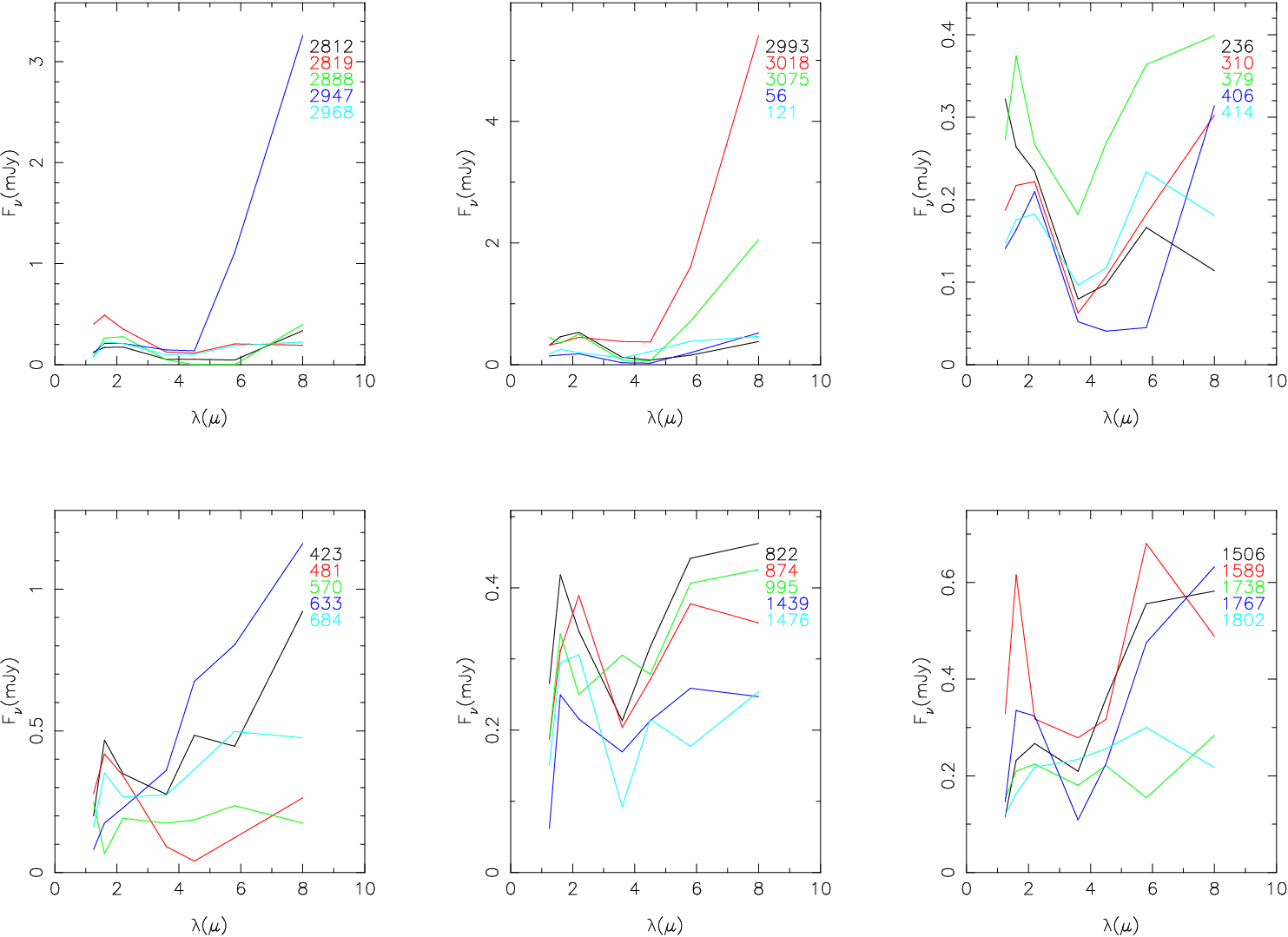}
\centerline{f12c.eps}
\end{figure}
\clearpage

\begin{figure}
\includegraphics[angle=-90,width=\textwidth]{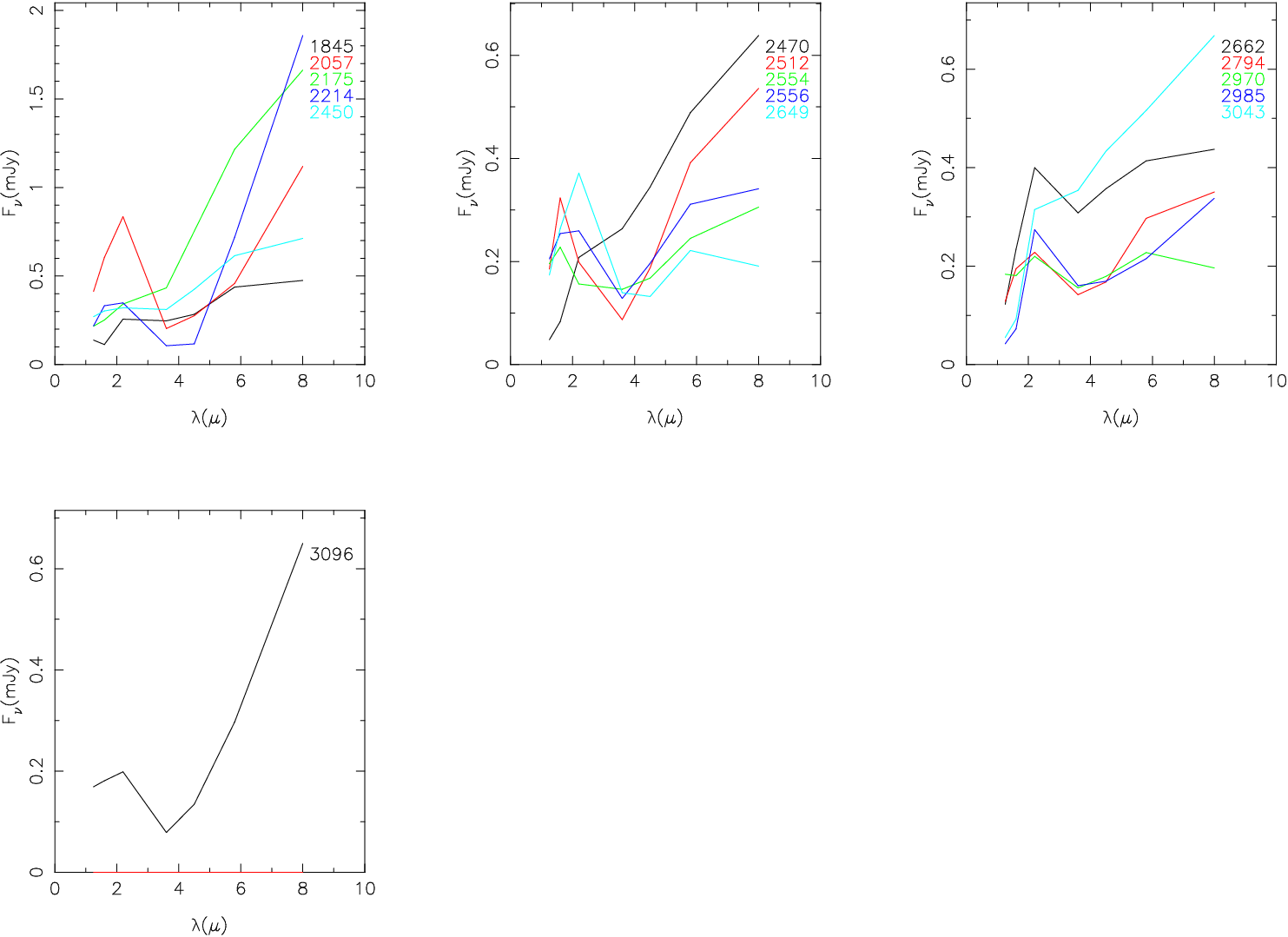}
\centerline{f12d.eps}
\end{figure}
\clearpage

\begin{figure}
\plotone{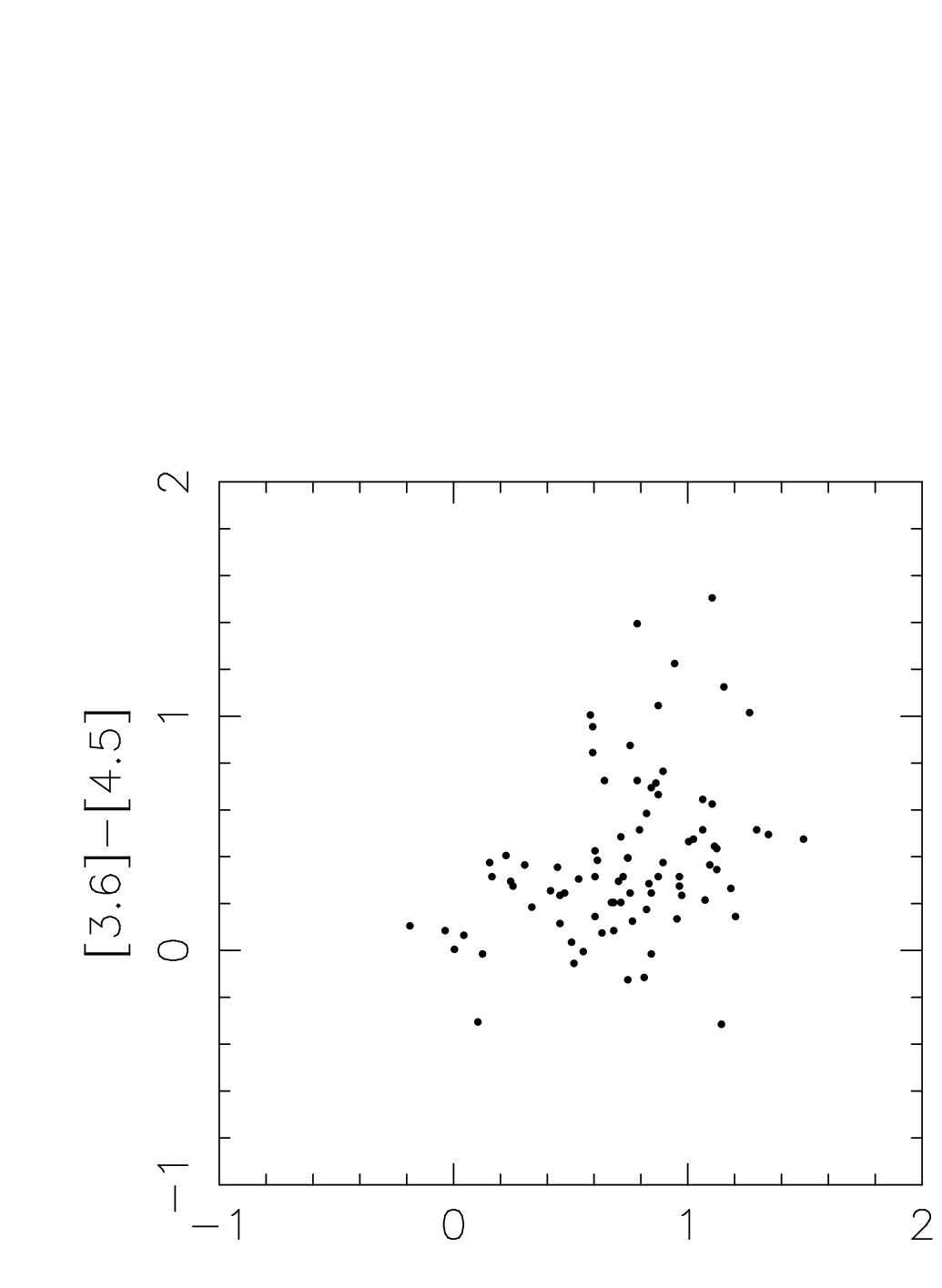}
\centerline{f13.eps}
\end{figure}
\clearpage

\begin{figure}
\includegraphics[angle=-0,width=0.8\textwidth]{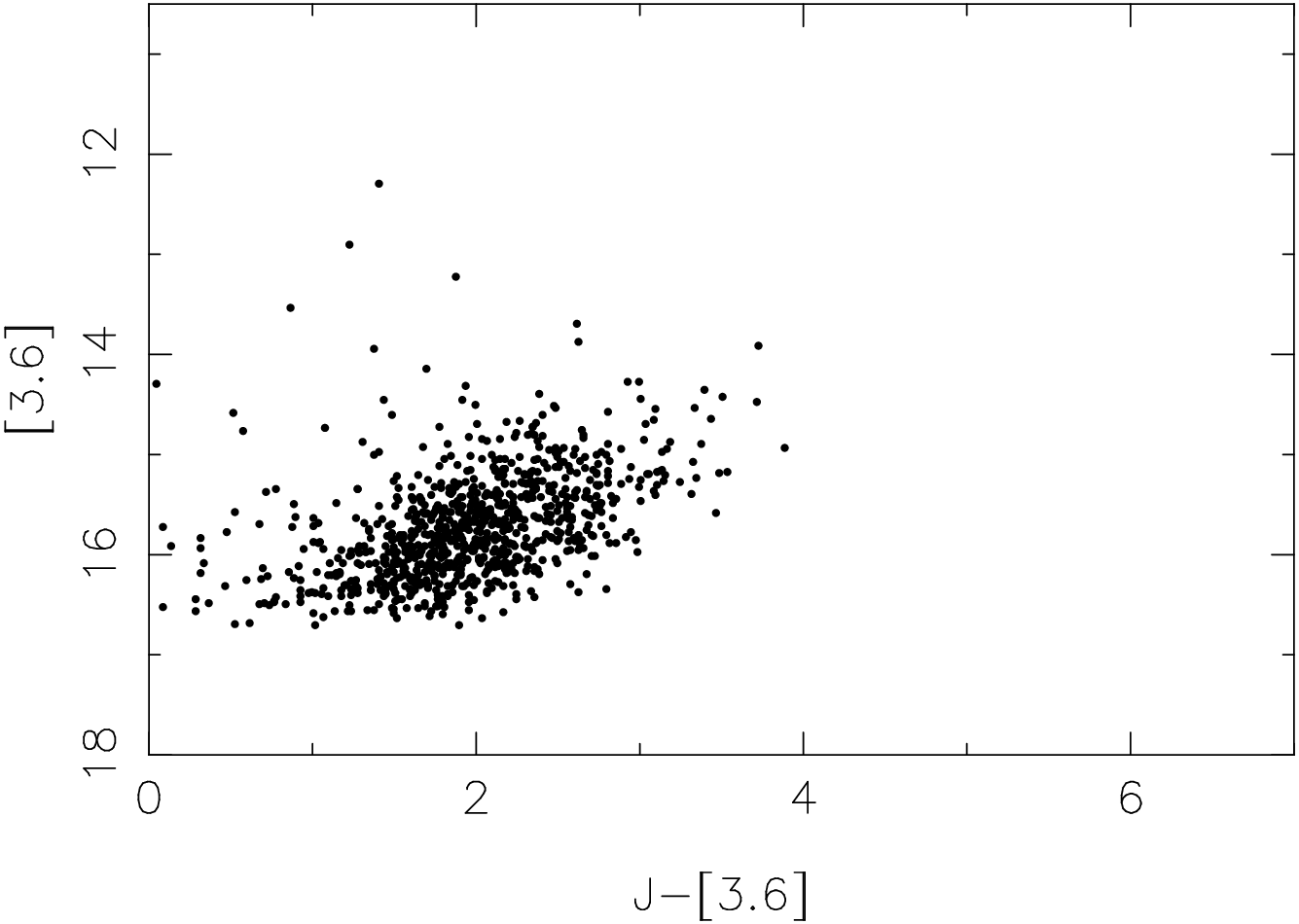}
\centerline{f14.eps}
\end{figure}
\clearpage

\begin{figure}
\plotone{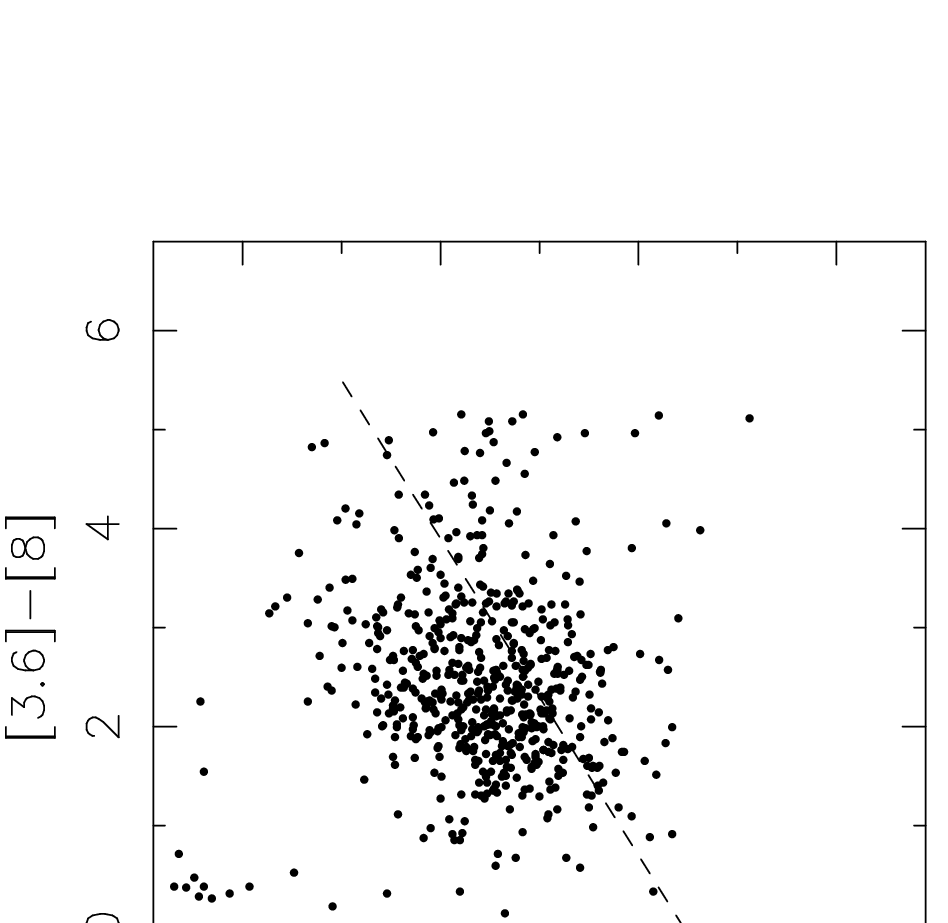}
\centerline{f15.eps}
\end{figure}
\clearpage

\begin{figure}
\includegraphics[angle=0,width=0.8\textwidth]{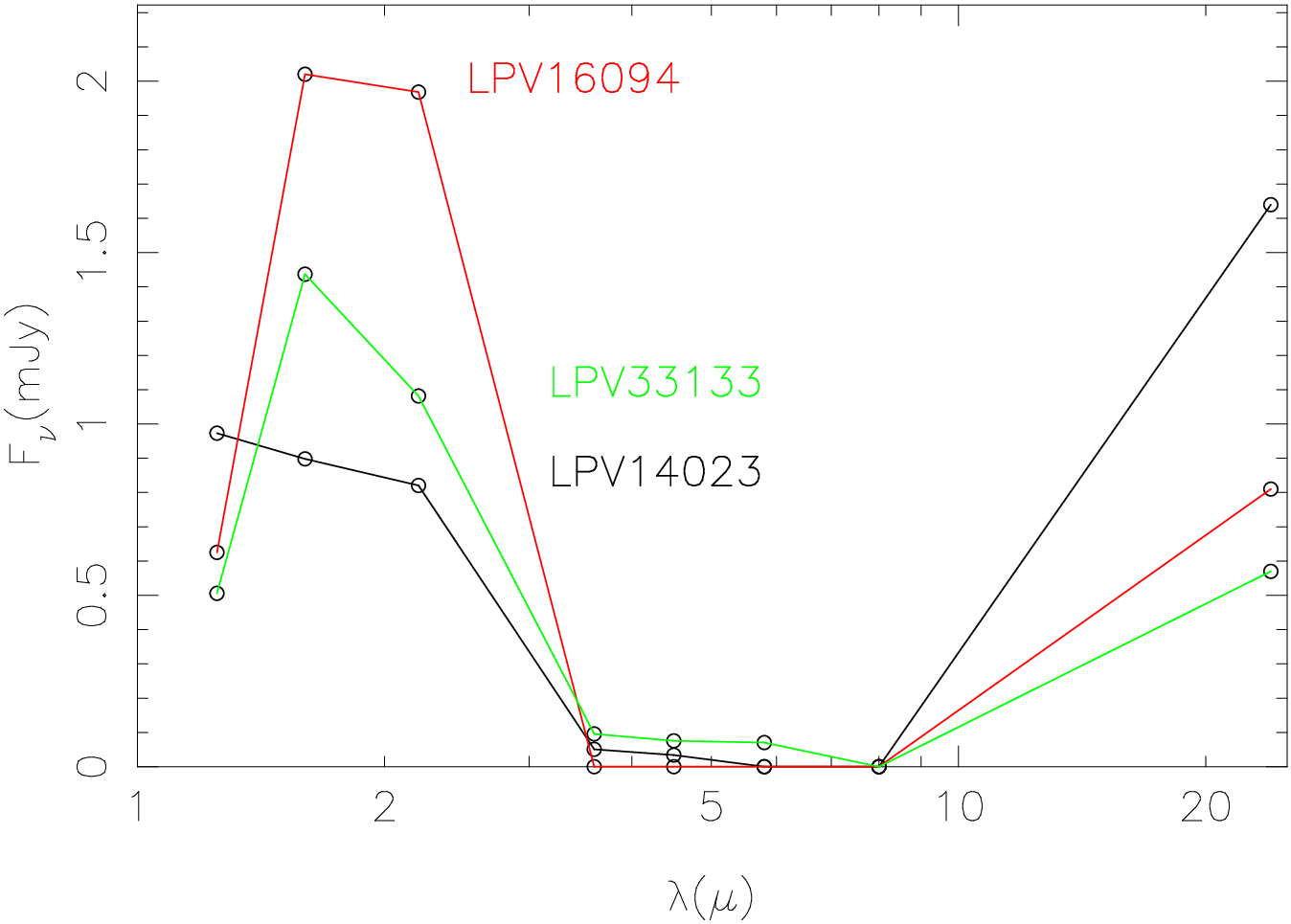}
\centerline{f16.eps}
\end{figure}
\clearpage

\begin{figure}
\includegraphics[angle=0,width=0.8\textwidth]{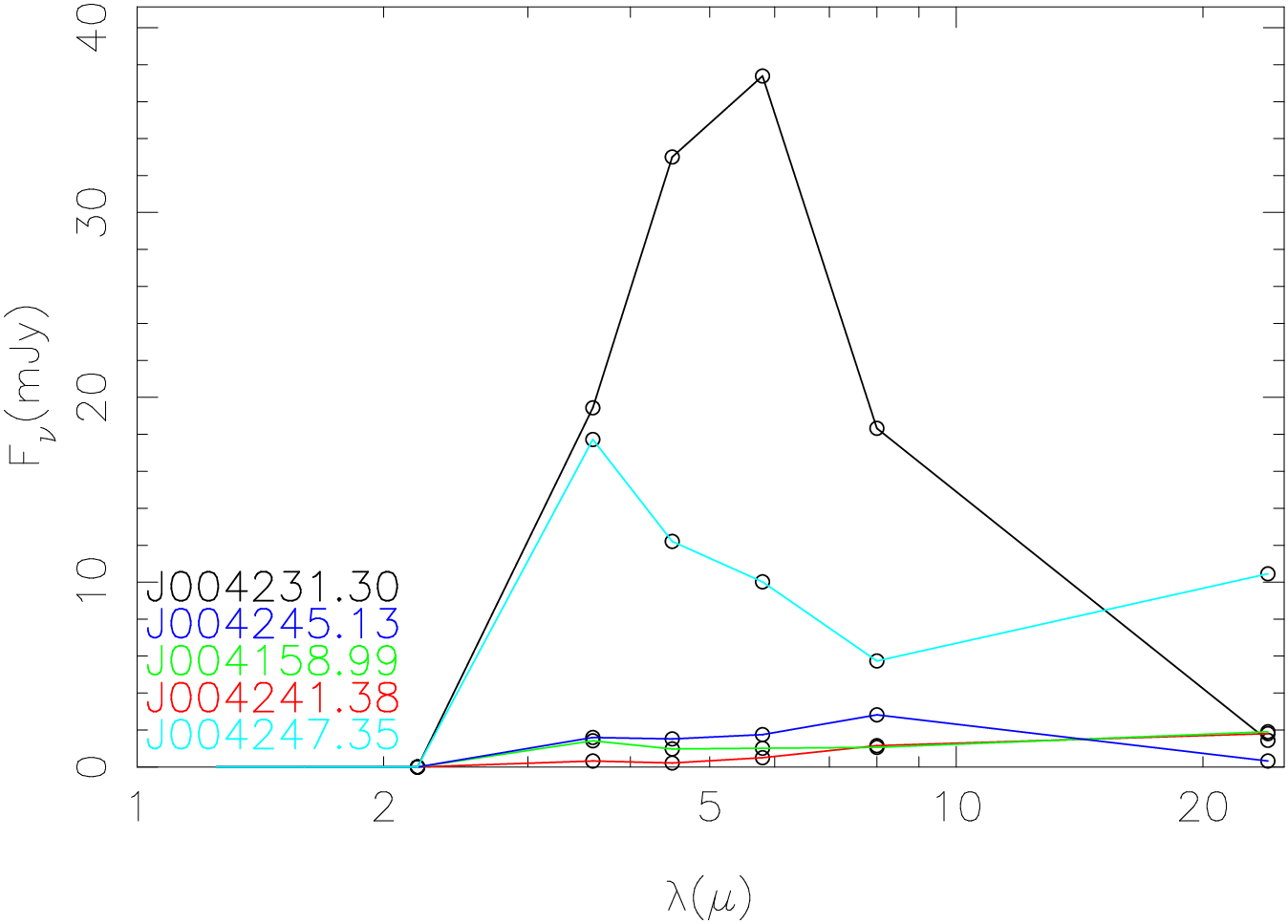}
\centerline{f17.eps}
\end{figure}
\clearpage

\begin{figure}
\includegraphics[angle=0,width=0.8\textwidth]{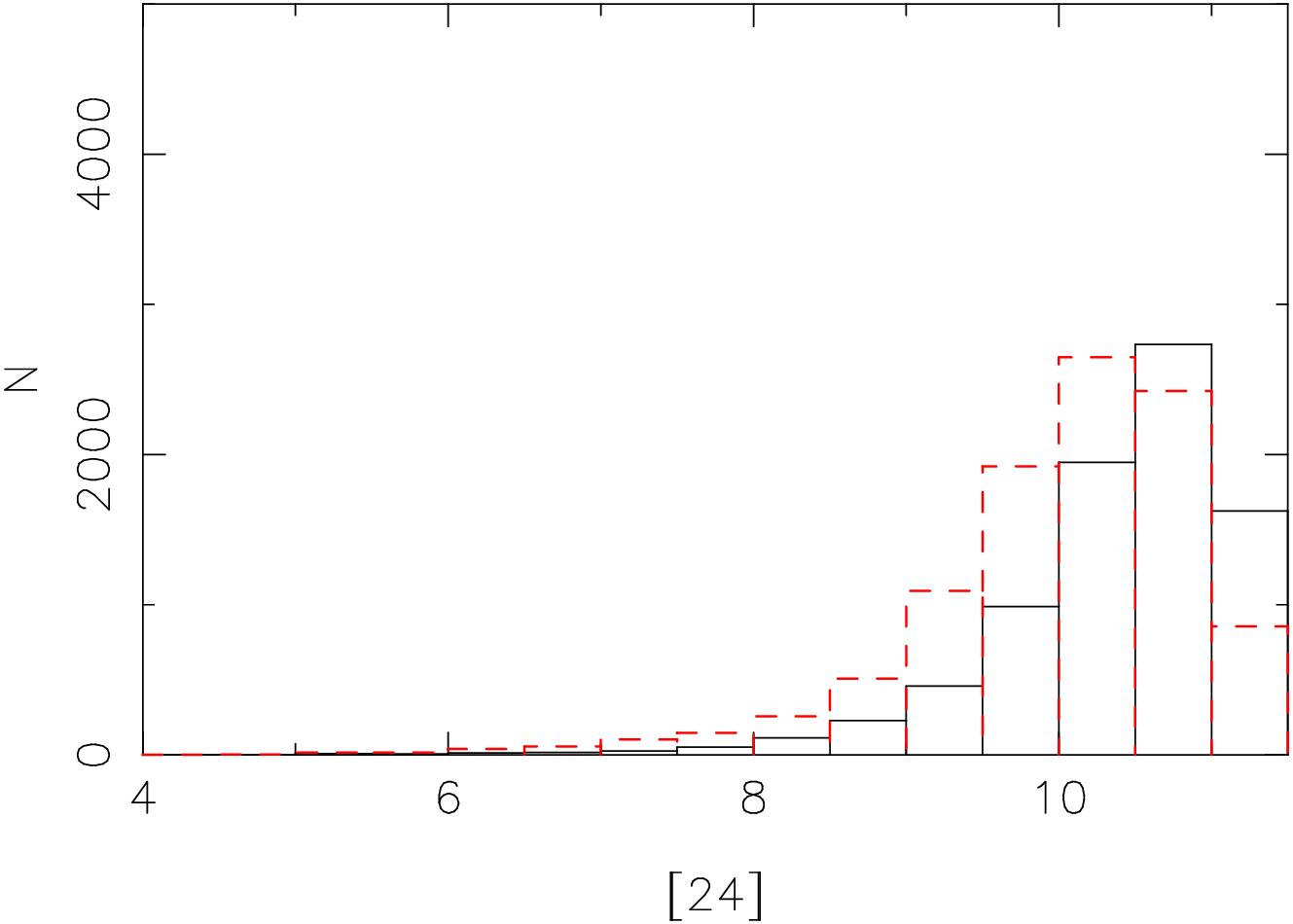}
\centerline{f18.eps}
\end{figure}

\end{document}